\documentclass[
superscriptaddress,
twocolumn,
showpacs,preprintnumbers,
amsmath,amssymb,
aps,
prb,
]{revtex4-1}

\usepackage{graphicx}% Include figure files
\usepackage{dcolumn}% Align table columns on decimal point
\usepackage{bm}% bold math
\usepackage{multirow}
\usepackage{float}
\usepackage{setspace}

\begin{document}
\title{Two-dimensional ferromagnetic semiconductor VBr$_{3}$ with tunable anisotropy}

% Force line breaks with \\
%\thanks{A footnote to the article title}%

\author{Lu Liu}
 \affiliation{Laboratory for Computational Physical Sciences (MOE),
 State Key Laboratory of Surface Physics, and Department of Physics,
 Fudan University, Shanghai 200433, China}

\author{Ke Yang}
 \affiliation{Laboratory for Computational Physical Sciences (MOE),
 State Key Laboratory of Surface Physics, and Department of Physics,
  Fudan University, Shanghai 200433, China}

\author{Guangyu Wang}
 \affiliation{Laboratory for Computational Physical Sciences (MOE),
 State Key Laboratory of Surface Physics, and Department of Physics,
  Fudan University, Shanghai 200433, China}

\author{Hua Wu}
\email{Corresponding author. wuh@fudan.edu.cn}
\affiliation{Laboratory for Computational Physical Sciences (MOE),
 State Key Laboratory of Surface Physics, and Department of Physics,
 Fudan University, Shanghai 200433, China}
\affiliation{Collaborative Innovation Center of Advanced Microstructures,
 Nanjing 210093, China}

%\collaboration{MUSO Collaboration}%\noaffiliation
%
%\author{Charlie Author}
% \homepage{http://www.Second.institution.edu/~Charlie.Author}
%\affiliation{
% Second institution and/or address\\
% This line break forced% with \\
%}%
%\affiliation{
% Third institution, the second for Charlie Author
%}%
%\author{Delta Author}
%\affiliation{%
% Authors' institution and/or address\\
% This line break forced with \textbackslash\textbackslash
%}%
%
%\collaboration{CLEO Collaboration}%\noaffiliation

%\date{\today}

\begin{abstract}
Two-dimensional (2D) ferromagnets (FMs) have attracted widespread attention due to their prospects in spintronic applications. Here we explore the electronic structure and magnetic properties of the bulk and monolayer of VBr$_{3}$ in the honeycomb lattice, using first-principles calculations, crystal field level analyses, and Monte Carlo simulations. Our results show that VBr$_{3}$ bulk has the $e'_{g}$$^2$ ($S$=1) ground state and possesses a small orbital moment and weak in-plane magnetic anisotropy. Those results well explain the recent experiments. More interestingly, we find that a tensile strain on the semiconducting VBr$_{3}$ monolayer tunes the ground state into $a_{1g}$$^1$$e'_{g}$$^1$ and thus produces a large orbital moment and a strong out-of-plane anisotropy. Then, the significantly enhanced FM superexchange and single ion anisotropy (SIA) would raise $T_{\rm C}$ from 20 K for the bare VBr$_{3}$ monolayer to 100-115 K under a 2.5$\%$-5$\%$ strain. Therefore, VBr$_{3}$ would be a promising 2D FM semiconductor with a tunable anisotropy.
\end{abstract}

\maketitle

\section{Introduction}
Bulk materials with a van der Waals (vdW) gap, readily to be cleaved, attract a large volume of attention due to their thickness-dependent electronic and magnetic properties.\cite{Huang_2017,Gong_2017,Deng_2018,Chen_2019,Li_2019,Song_2019,Klein_2019,Huang_2020,Kim_2019} In principle, long-range magnetic order at finite temperature is prohibited in two-dimensional (2D) isotropic Heisenberg spin systems according to the Mermin-Wagner theorem.\cite{MW} Very recently, 2D ferromagnetism (FM) has been observed in atomically thin CrI$_{3}$ (Ref. 1) and Cr$_{2}$Ge$_{2}$Te$_{6}$ (Ref. 2). The FM ordering is remarkably retained in CrI$_{3}$ monolayer with the Curie temperature $T_{\rm C}$=45 K. These discoveries have brought about intriguing magnetism and have invoked extensive research in 2D FMs. This opens a new avenue to spintronic applications, such as spin valves,\cite{valves} spin filters,\cite{filter1,filter2} and data storage.\cite{storage}

Vanadium trihalides are vdW materials with a layered honeycomb structure, see Fig. 1. They are of current interest due to their similarity with CrI$_{3}$, and they may also be promising candidates for 2D FMs.\cite{Huang_2017,Kong_VBr,Kong_2019,Tian_2019_jacs,Son_2019} It is worth noting that FM CrI$_{3}$, with a closed Cr$^{3+}$ $t_{2g}$$^{3}$ shell, has a quenched orbital moment, and that its magnetic anisotropy comes from an exchange anisotropy induced by the spin-orbit coupling (SOC) of the heavy I $5p$ orbitals and their hybridization with Cr $3d$.\cite{Lado_2017,Kim_2019_PRL} In contrast, the open V$^{3+}$ $t_{2g}$$^{2}$ shell in vanadium trihalides may carry an unquenched orbital moment and achieve a single ion anisotropy (SIA) via the V$^{3+}$ SOC. This seems to account for the hard perpendicular FM with $T_{\rm C}$$\approx$50 K observed in VI$_3$ bulk\cite{Kong_2019,Tian_2019_jacs,Son_2019} and even the Ising FM in VI$_3$ monolayer.\cite{Yang_2019}

Among vanadium trihalides, VBr$_{3}$ bulk is a layered antiferromagnetic (AF) semiconductor with the Neel temperature $T_{\rm N}$=26.5 K, the effective magnetic moment of 2.6 $\mu_{\rm B}$ per V$^{3+}$ ion, and the optical band gap slightly larger than 1 eV. And it has a stronger in-plane magnetic susceptibility than the out-of-plane one below $T_{\rm N}$.\cite{Kong_VBr} In this article, we study the electronic and magnetic structures of VBr$_{3}$ bulk and monolayer from first-principles calculations, crystal field level analyses, and Monte Carlo simulations. We find that the V$^{3+}$ ions are in the $e'_{g}$$^2$ $S$=1 state in the trigonal crystal field of the honeycomb lattice, and they have an intralayer FM coupling but one order of magnitude weaker interlayer magnetic coupling. Moreover, each V$^{3+}$ ion carries a small in-plane orbital moment due to a mixing of the $e'_{g}$ and $a_{1g}$ states by V$^{3+}$ SOC. Then, VBr$_{3}$ bulk would rather have a weak parallel magnetic anisotropy as detailed below, and this explains the experimental anisotropic magnetic susceptibilities. Note that VBr$_3$ monolayer, if cleaved, would be a 2D FM semiconductor with a weak parallel magnetic anisotropy and $T_{\rm C}$$\approx$20 K. More interestingly, VBr$_3$ monolayer would turn into the $a_{1g}$$^1$$e'_{g}$$^1$ ground state under a tensile strain. Then it carries a large out-of-plane orbital moment and has a strong SIA driven perpendicular magnetic anisotropy. As a result, the significantly enhanced FM superexchange and perpendicular anisotropy would raise $T_{\rm C}$ up to 100-115 K under a 2.5$\%$-5$\%$ strain. Therefore, VBr$_{3}$ monolayer would be an appealing 2D FM semiconductor with a tunable anisotropy.

\begin{figure}[t]
  \includegraphics[width=9cm]{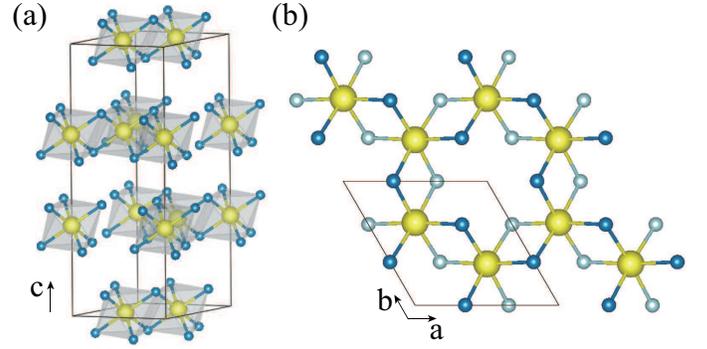}
  \caption{(a) The crystal structure of VBr$_{3}$ bulk with V (Br) atoms represented by yellow (blue) balls. (b) The structure of VBr$_{3}$ monolayer, and the blue (grey) balls referring to the upper (lower) Br atoms within the monolayer.}
  \label{fgr:example}
\end{figure}

\section{Computational Details}

VBr$_{3}$ bulk material (space group R$\overline{3}$) has the A-B-C layer stacking sequence via vdW interactions, see Fig. 1. Edge-sharing VBr$_{6}$ octahedra form a honeycomb lattice within each layer. For V$^{3+}$ ions with the local octahedral coordinates, the global trigonal crystal field splits the $t_{2g}$ triplet into the $a_{1g}$ singlet and the $e'_{g}$ doublet. The global coordinate system was used in the following calculations, with the $z$ axis along the $\left[ 111 \right]$ direction of the local VBr$_{6}$ octahedra and $y$ along the $\left[ 1\overline{1}0 \right]$ direction. Then the $t_{2g}$ wave functions under the trigonal crystal field can be written as
\begin{equation}
\begin{aligned}
&\left|a_{1 g}\right\rangle=\left|3 z^{2}-r^{2}\right\rangle \\
&\left|e_{g 1}^{\prime}\right\rangle= \sqrt{\frac{2}{3}}\left|x^{2}-y^{2}\right\rangle-\sqrt{\frac{1}{3}}\left|xz\right\rangle \\
&\left|e_{g 2}^{\prime}\right\rangle= \sqrt{\frac{2}{3}}\left|xy\right\rangle+\sqrt{\frac{1}{3}}\left|yz\right\rangle. \\
\end{aligned}
\end{equation}

Density functional calculations (DFT) were carried out using the full-potential augmented plane wave plus local orbital code (Wien2k).\cite{WIEN2K} The lattice parameters of VBr$_{3}$ bulk (monolayer) were optimized to be $a$=$b$=6.299 \AA~(6.342 \AA) and $c$=18.110 \AA, which are almost the same (within 1.5\%) as the experimental ones of $a$=$b$=6.371 \AA~and $c$=18.376 \AA.\cite{Kong_VBr} A vacuum slab of 10 \AA~was set along the $c$-axis for the monolayer. The muffin-tin sphere radii were chosen to be 2.5 Bohr for V atoms and 2.2 Bohr for Br. The plane-wave cut-off energy of 12 Ry was set for the interstitial wave functions, and a $12\times12\times1$ $k$-mesh was used for integration over the Brillouin zone. Note that VBr$_{3}$ is a narrow band (less than 1 eV, see below) strongly correlated system, and the electron correlation was included in our calculations, using the local spin density approximation plus Hubbard $U$ (LSDA+$U$) method,\cite{Anisimov_1993} with the typical values of Hubbard $U$=3.9 eV and Hund exchange $J\rm _{H}$=0.9 eV. The obtained semiconducting solution with the band gap of 1.2 eV well reproduces the experimental band gap. We also use crystal-field level diagrams and superexchange pictures to understand the electronic and magnetic properties. The spin-orbit coupling (SOC) was included for V $3d$ and Br $4p$ orbitals by the second-variational method with scalar relativistic wave functions. To clearly see the crystal field effect, exchange splitting, electron correlation, and the crucial SOC effects, we present and discuss below the spin restricted LDA, spin polarized LSDA, LSDA+$U$, and LSDA+SOC+$U$ calculations. Monte Carlo simulations on a $6\times6\times1$ spin matrix have also been performed to estimate the Curie temperature of VBr$_{3}$ monolayer using Metropolis method.\cite{Nicholas_1949}

\section{Results and Discussion}
\begin{center}
\textbf{VBr$_{3}$ Bulk}
\end{center}

We first investigate the VBr$_{3}$ bulk for which the experimental results are available for comparison. Our LDA calculations find the $t_{2g}$-$e_g$ crystal field splitting of about 1.5 eV, see Fig. 2(a). The $a_{1g}$ singlet and $e'_g$ doublet out of $t_{2g}^2$ each forms a partially occupied narrow band (less than 1 eV) crossing Fermi level and they are almost degenerate. The Br $4p$ state lies in the range of 2-6 eV below Fermi level, and it has a notable $pd$ hybridization with V $3d$, particularly the strong $pd\sigma$ one with the $e_g$ orbital. Owing to the narrow band localization effect, the V $3d$ electrons prefer to be spin polarized and they form $S$=1 state as indicated by LSDA calculations. As seen in Fig. 2(b), only the up-spin $a_{1g}$ and $e'_g$ bands are partially occupied, giving the total spin moment of 2 $\mu_{\rm B}$ for each V$^{3+}$ ion.

 \begin{figure}[t]
\centering
\includegraphics[scale=0.7]{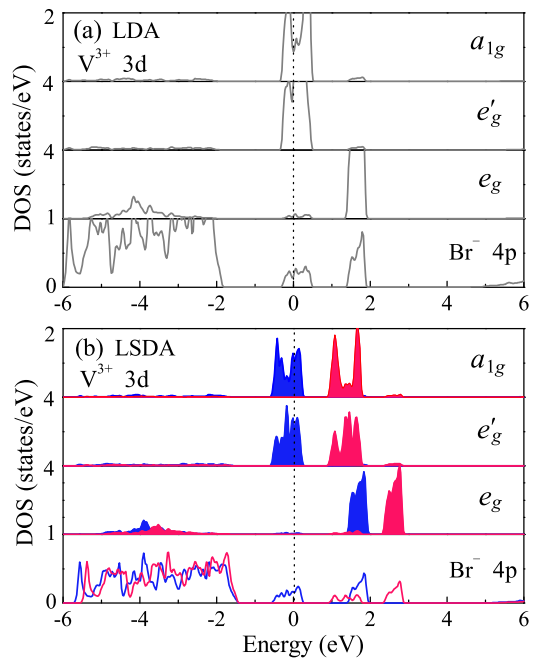}
 \caption{The DOS results of (a) LDA and (b) LSDA for bulk VBr$_{3}$ in the trigonal crystal field.  The blue (red) curves refer to majority (minority) spins. The Fermi level is set at the zero energy.}
\label{fig1:el}
\end{figure}

Moreover, when the electron correlation effect is included for this narrow band system as done in our LSDA+$U$ calculations, a semiconducting solution is obtained with the closed $e'_{g}$$^2$ subshell of the V$^{3+}$ ions, see Fig. 3(a). The semiconducting gap of 1.2 eV is well comparable with the experimental value slightly larger than 1 eV. As seen in Table I, the $S$=1 V$^{3+}$ ion has the local spin moment of 1.86 $\mu_{\rm B}$ reduced by the V $3d$-Br $4p$ hybridization. In contrast, another metallic solution ($a_{1g}$$^1$$e'_g$$^1$), with the half-filled $e'_g$ band crossing Fermi level, was also obtained in our LSDA+$U$ calculations, see Fig. 3(b). However, this metallic solution turns out to be much more unstable against the above semiconducting one by about 312.5 meV/fu, see Table I. Therefore, the V$^{3+}$ $e'_g$$^2$ solution seems to be the ground state of VBr$_{3}$ bulk. For this orbital state, our LSDA+$U$ calculations find that the intralayer FM state is more stable than the AF state by 7.9 meV/fu, but the interlayer magnetic coupling is one order of magnitude weaker (see below) and is out of concern (and beyond the scope) of this work.

To analyze the above results, we plot in Fig. 4 the crystal field level diagrams of the V$^{3+}$ ions. As the $a_{1g}$ singlet and $e'_g$ doublet are almost degenerate, the two possible solutions, $e'_g$$^2$ and $a_{1g}$$^1$$e'_g$$^1$ are displayed in Figs. 4(a) and 4(d), respectively. The former is semiconducting but the latter is metallic, given by the above LSDA+$U$ calculations. However, when the SOC is included, the $a_{1g}$ and $e'_g$ can be mixed. Then the SOC can produce an in-plane orbital moment, for example, an orbital moment along the $x$ axis by mixing the $a_{1g}$ and $e'_{g2}$ [see Eq. 1 and Figs. 4(b)-4(c)]. In contrast, the mixing between the $e'_{g1}$ and $e'_{g2}$ states by the SOC will produce the $L_{z\pm}$ states with an orbital moment along the $z$ axis. These possible solutions, as shown in Figs. 4(b), 4(c), 4(e), and 4(f), will be obtained in the following LSDA+SOC+$U$ calculations.

 \begin{figure}[t]
 \centering
\includegraphics[scale=0.7]{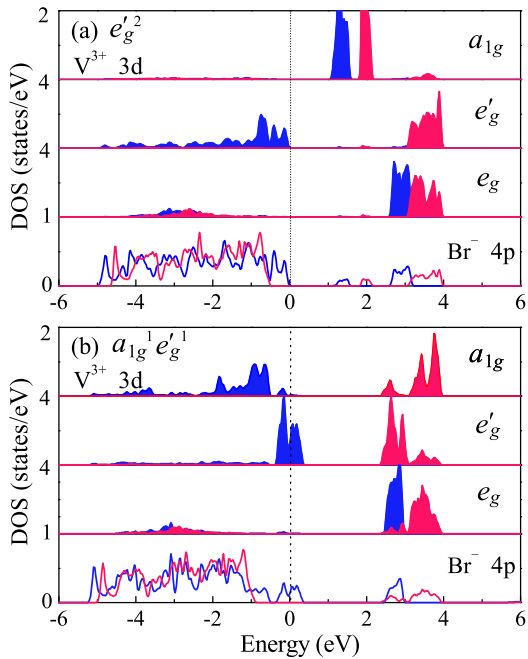}
 \caption{The DOS results of LSDA+$U$ for the insulating $e'_{g}$$^2$ state (a) and the metallic $a_{1g}$$^1$$e'_{g}$$^1$ state (b). The blue (red) curves refer to majority (minority) spins. The Fermi level is set at the zero energy.}
\label{fig1:el}
\end{figure}
\renewcommand\arraystretch{1.5}
\begin{table}[t]
  \caption{Relative total energies $\Delta E$ (meV/fu) for VBr$_{3}$ bulk by the LSDA+$U$ and LSDA+SOC+$U$ calculations, and the local spin and orbital moments ($\mu_{\rm B}$) for the V$^{3+}$ ion. The FM state is considered in most calculations except for those marked with intralayer AF, and $\parallel$ ($\perp$) represents the in-plane (out-of-plane) magnetization.
}
  \label{tb1}
\setlength{\tabcolsep}{2mm}{
\begin{tabular*}{0.48\textwidth}{@{\extracolsep{\fill}}llrrrr}
\hline\hline
  & States   & $\Delta E$ & $M_{\rm spin}$    & $M_{\rm orb}$ \\ \hline
LSDA+$U$ &$e'_{g}$$^{2}$      & 0.0  & 1.86      & $-$          \\
&$e'_{g}$$^{2}$ (AF)      & 7.9  & 1.84       & $-$     \\
 &$a_{1g}$$^{1}$$e'_{g}$$^{1}$       & 312.5  & 2.04 & $-$ \\\hline
LSDA+SOC+$U$ &$e'_{g}$$^{2}$, $\parallel$   & 0.0  & 1.87   & $-$0.23 \\
&$e'_{g}$$^{2}$, $\parallel$ (AF)   & 8.2      & 1.84      &  $-$0.21 \\
& $e_{g}^{\prime}$$^{2}$, $\perp$   & 2.3      & 1.86            & 0.00   \\
& $a_{1g}$$^{1}$$L_{z-}$$^{1}$, $\perp$   & 14.5      & 1.91     & $-$1.15   \\
& $e'_{g1}$$^{1}$$L_{x-}$$^{1}$, $\parallel$  & 26.7   & 1.90   & $-$1.11    \\
& $a_{1g}$$^{1}$$L_{z+}$$^{1}$, $\perp$   & 54.1    & 1.91    & 1.14   \\
\hline\hline
 \end{tabular*}}
\end{table}
 \begin{figure}[b]
  \centering
\includegraphics[width=9cm]{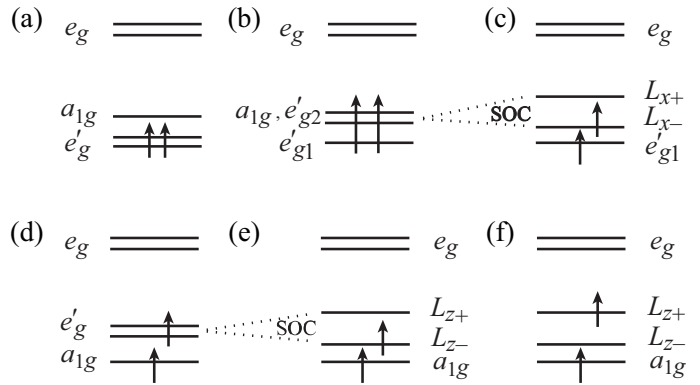}
 \caption{Crystal field level diagrams for V$^{3+}$ $S$=1 ion in different configuration states: (a) $e'_{g}$$^2$, (b) $e'_{g1}$$^1$$(a_{1g}e'_{g2})$$^1$, and (d) $a_{1g}$$^1$$e'_{g}$$^1$. SOC is active in (c), (e) and (f).
 }
\label{fig1:el}
\end{figure}
 \begin{figure}[t]
  \centering
\includegraphics[scale=0.7]{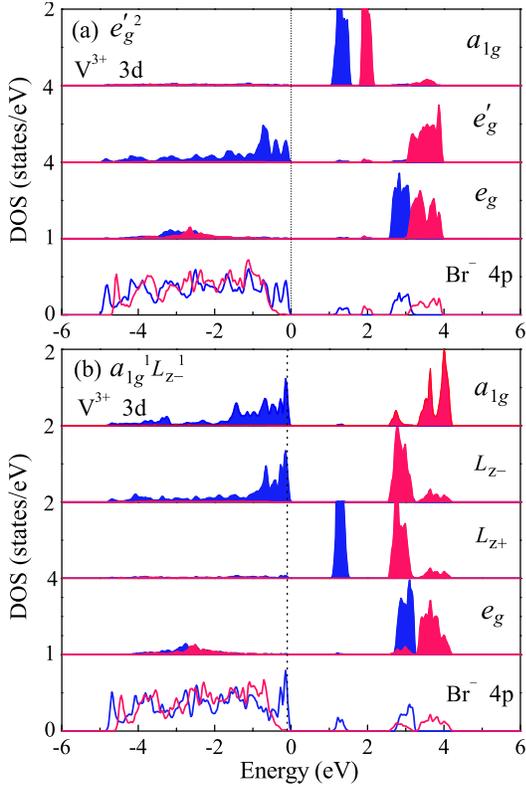}
 \caption{The DOS results of the $e'_{g}$$^2$ ground state (a) and the $a_{1g}$$^{1}$$L_{z-}$$^{1}$ meta-stable state (b) of bulk VBr$_{3}$ by LSDA+SOC+$U$. The blue (red) curves stand for majority (minority) spins. The Fermi level is set at the zero energy.
 }
\label{fig1:el}
\end{figure}

We collect our LSDA+SOC+$U$ results in Table I. The formal $e'_g$$^2$ state now has an in-plane orbital moment of --0.23 $\mu_{\rm B}$, in addition to the spin moment of 1.87 $\mu_{\rm B}$. This semiconducting solution is very similar to that one by LSDA+$U$, see Figs. 3(a) and 5(a). The small in-plane orbital moment tends to align via SOC the spin moment also in the plane. If the magnetization is assumed to be along the $z$ axis for the formal $e'_g$$^2$ state, no orbital moment will appear as the SOC split $L_{z\pm}$ states are both fully occupied. Then only the spin moment of 1.86 $\mu_{\rm B}$ persists. As a result, the formal $e'_g$$^2$ state with the small in-plane orbital moment gains a partial SOC energy, producing the easy planar magnetization with the magnetic anisotropy energy of 2.3 meV/fu, see Table I. Moreover, this state favors the intralayer FM coupling, with the intralayer AF state being higher in energy by 8.2 meV/fu. This intralayer FM coupling will be further elucidated below for the VBr$_{3}$ monolayer. In contrast, the interlayer magnetic coupling is much weaker due to the vdW gap and here the layered FM and AF states differ in total energy only by 0.8 meV/fu.

To determine the electronic ground state of the VBr$_{3}$ bulk, we also achieve several other possible solutions using the constrained LSDA+SOC+$U$ calculations with the initialized configurations~\cite{Ou_2014,Ou_2015} for those different orbital states shown in Fig. 4. For example, when we start from the $a_{1g}$$^1$$e'_g$$^1$ state and now perform LSDA+SOC+$U$ calculations, we will get either $a_{1g}$$^1$$L_{z-}$$^1$ state or $a_{1g}$$^1$$L_{z+}$$^1$. The former solution is now semiconducting [see Fig. 5(b)], in sharp contrast to the $a_{1g}$$^1$$e'_g$$^1$ metallic one given by LSDA+$U$ [see Fig. 3(b)]. This semiconducting $a_{1g}$$^1$$L_{z-}$$^1$ state lies in total energy higher than the ($e'_{g}$$^{2}$ $\parallel$) ground state by 14.5 meV/fu, and has the local spin moment of 1.91 $\mu_{\rm B}$ and the orbital moment of --1.15 $\mu_{\rm B}$ (see Table I), both of which are along the $z$ axis but antiparallel. If we force the orbital moment to be parallel to the spin moment as done for the $a_{1g}$$^1$$L_{z+}$$^1$ solution, the increasing total energy (54.1 vs 14.5 meV/fu) allows us to estimate the SOC strength parameter $\xi\sim$ 40 meV for V$^{3+}$ $3d$ electrons. Moreover, by an equal mixing via SOC between the $a_{1g}$ and $e'_{g2}$ states, the $e'_{g1}$$^1$$L_{x-}$$^1$ state can also be obtained in our LSDA+SOC+$U$ calculations, which give a spin (orbital) moment of 1.90 (--1.11) $\mu_{\rm B}$ both along the $x$ axis. However, this solution is higher in total energy than the ($e'_{g}$$^{2}$ $\parallel$) ground state by 26.7 meV/fu.

The above results lead us to a conclusion that the vdW bulk material VBr$_{3}$ is a narrow band magnetic semiconductor. It is in the formal V$^{3+}$ $e'_g$$^2$ $S$=1 ground state and has a small in-plane orbital moment and magnetic anisotropy energy of 2.3 meV/fu. The spin and orbital moments prefer to align in the plane, and those V$^{3+}$ spins are FM coupled in the plane. The V$^{3+}$ $S$=1 spin moment and a small antiparallel orbital moment account for the experimental effective magnetic moment of 2.6 $\mu_{\rm B}$,~\cite{Kong_VBr} which is slightly reduced from the pure spin contribution 2$\sqrt{S(S+1)}$=2.83 $\mu_{\rm B}$. Moreover, considering the intralayer FM coupling and the one order of magnitude weaker interlayer magnetic coupling and the experimental layered AF state with $T_{\rm N}$=26.5 K, we can explain the experimentally observed magnetic anisotropy~\cite{Kong_VBr}: At 50 K and above, the magnetic susceptibilities are almost the same for the applied magnetic field parallel or perpendicular to the $ab$ plane; At 1.8 K, the in-plane magnetic susceptibility is much stronger than the out-of-plane one. This is because at 1.8 K, VBr$_{3}$ bulk is in the layered AF state with the spins lying in the $ab$ plane. As the interlayer AF coupling ($\sim$0.8 meV/fu) is weaker than the magnetic anisotropy (2.3 meV/fu), those spins are more susceptible to the in-plane magnetic field and can flip in the $ab$ plane, giving a stronger parallel magnetic susceptibility. At 50 K and above, the magnetic anisotropy (2.3 meV) is already overcome, and thus the system is in the isotropic paramagnetic state. As such, our results have well explained the experimental magnetic behavior of VBr$_{3}$ bulk.

\subsection*{VBr$_{3}$ Monolayer}
 \begin{figure}[b]
  \centering
\includegraphics[width=7cm,height=5cm]{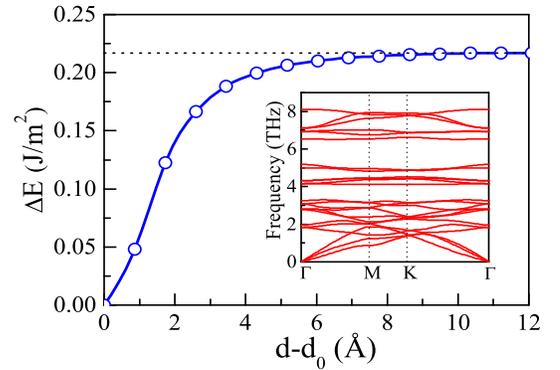}
 \caption{The relative energy as a function of the distance between two VBr$_{3}$ monolayers with respect to the experimental vdW distance d$_{0}$, and the phonon spectrum shown in the inset.
 }
\label{fig1:el}
\end{figure}

The vdW material VBr$_{3}$ has an intralayer FM coupling and a much weaker interlayer magnetic coupling, and thus it could be a potential 2D FM semiconductor which is desirable for spintronics. The calculated cleavage energy of 0.22 J/m$^2$ for the VBr$_{3}$ monolayer (see Fig. 6), using the DFT+vdW correction,\cite{vdw} is well comparable with that of 0.3 J/m$^2$ for the CrI$_3$ monolayer which is already exfoliated from its bulk. The dynamical stability of VBr$_{3}$ monolayer is also indicated by our phonon calculations,~\cite{vasp,phonopy} which show no imaginary frequency phonons through the whole Brillouin zone. Hence, we now turn to the VBr$_{3}$ monolayer and focus on its electronic and magnetic properties by carrying out LSDA+SOC+$U$ calculations, superexchange analyses, and Monte Carlo simulations of its 2D FM order. The tensile strain effect of the 2D lattice is also studied, and it is capable of tuning the orbital state of the V$^{3+}$ ions and turns out to significantly enhance the FM order with tunable superexchange and magnetic anisotropy.

The electronic structure of the VBr$_{3}$ monolayer is very similar to that of the bulk, and therefore the DOS results for the monolayer are not shown here again. We list the major results in Table II from the LSDA+SOC+$U$ calculations of the most concern. All the solutions for VBr$_{3}$ monolayer are semiconducting. For the bare monolayer without strain, the electronic ground state is again the formal $e'_{g}$$^2$ state, which carries the local V$^{3+}$ spin moment of 1.87 $\mu_{\rm B}$ and a small in-plane orbital moment of --0.23 $\mu_{\rm B}$.
As a result, the easy magnetization direction lies in the $ab$ plane, with the magnetic anisotropy energy of 2.0 meV/fu. Moreover, this $e'_g$$^2$ ground state prefers the FM coupling within the 2D V$^{3+}$ lattice, which is more stable than the AF state by 7.9 meV/fu. In addition, for the $a_{1g}$$^1$$L_{z-}$$^1$ state, it has a large out-of-plane orbital moment of --1.15 $\mu_{\rm B}$ (antiparallel to the spin moment of 1.91 $\mu_{\rm B}$) but is less stable than the $e'_g$$^2$ FM ground state by 12.0 meV/fu. Therefore, the bare VBr$_3$ monolayer is a FM semiconductor with a parallel magnetic anisotropy.
\renewcommand\arraystretch{1.5}
\begin{table}[t]
% \scriptsize
%\footnotesize
%\tiny
  \caption{Relative total energies $\Delta E$ (meV/fu), local spin and orbital moments ($\mu_{\rm B}$/V) of VBr$_{3}$ monolayer by LSDA+SOC+$U$. The FM state is considered in most cases except for the states marked with AF, and $\parallel$ ($\perp$) stands for the in-plane (out-of-plane) magnetization.
}
\label{tb1}
\setlength{\tabcolsep}{1mm}{
\begin{tabular*}{0.48\textwidth}{@{\extracolsep{\fill}}llrrrr}
\hline\hline
Strain  & States &$\Delta E$ & $M_{\rm spin}$   & $M_{\rm orb}$ \\\hline
0\% &  $e'_{g}$$^{2}$, $\parallel$ ~~  & 0.0      & 1.87             & $-$0.23    \\
& $e'_{g}$$^{2}$, $\parallel$ (AF)   & 7.9      & 1.84      &  $-$0.21     \\
& $e_{g}^{\prime}$$^{2}$, $\perp$  & 2.0      & 1.86            & 0.00   \\
& $e_{g}^{\prime}$$^{2}$, $\perp$ (AF)  & 9.7      & 1.84            & 0.00   \\
& $a_{1g}$$^{1}$$L_{z-}$$^{1}$, $\perp$~~   & 12.0      & 1.91     & $-$1.15   \\\hline
2.5\% & $a_{1g}$$^{1}$$L_{z-}$$^{1}$, $\perp$~~  & 0.0     & 1.92    & $-$1.16  \\
& $a_{1g}$$^{1}$$L_{z-}$$^{1}$, $\perp$ (AF)  & 31.8     & 1.87    & $-$1.22  \\
& $a_{1g}$$^{1}$$L_{z-}$$^{1}$, $\parallel$   & 15.0     & 1.92    & $-$0.12  \\
& $a_{1g}$$^{1}$$L_{z-}$$^{1}$, $\parallel$ (AF)  & 47.3     & 1.87    & $-$0.10  \\
& $e'_{g}$$^{2}$, $\parallel$   & 80.9      & 1.87     &  $-$0.44     \\\hline
5\% & $a_{1g}$$^{1}$$L_{z-}$$^{1}$, $\perp$~~  & 0.0     & 1.92    & $-$1.17  \\
& $a_{1g}$$^{1}$$L_{z-}$$^{1}$, $\perp$ (AF)  & 38.4     & 1.87    & $-$1.25  \\
& $a_{1g}$$^{1}$$L_{z-}$$^{1}$, $\parallel$   & 13.9     & 1.93    & $-$0.09  \\
& $a_{1g}$$^{1}$$L_{z-}$$^{1}$, $\parallel$ (AF)  & 52.4     & 1.87    & $-$0.08  \\
& $e'_{g}$$^{2}$, $\parallel$   & 141.6      & 1.87     &  $-$0.41     \\\hline\hline
Strain  &~~~~~~~~~~~~$D$   &$J$~~~~~~  &$J^{\prime}$~~  &$T_{\rm C}$ \\\hline
0\%, $M\parallel$  &~~~~~~~~$-$1.9  &2.6~~~~~~  &$-0.06$~~  &~~~20\\
2.5\%, $M\perp$  &~~~~~~~~~15.2  &10.8~~~~~~  &$-0.15$~~  &~~~100\\
5\%, $M\perp$  &~~~~~~~~~13.9  &12.8~~~~~~  &$-0.02$~~  &~~~115\\\hline\hline
\end{tabular*}}
\end{table}

The electronic state and magnetism of 2D materials may be tuned by a strain.~\cite{ref1,ref2,ref3} For VBr$_3$ monolayer, a tensile strain would flatten the VBr$_6$ octahedron along the $z$ axis (i.e., the local [111] direction of the VBr$_3$ octahedron) and lowers the $a_{1g}$ singlet off the $e'_g$ doublet. Then, the $a_{1g}$$^1$$e'_g$$^1$ would become the ground state for the $S$=1 V$^{3+}$ ion, and it would have a large perpendicular orbital moment and a strong single ion anisotropy (SIA) in the SOC split ground state $a_{1g}$$^1$$L_{z-}$$^1$. To verify this, we perform LSDA+SOC+$U$ calculations for the VBr$_3$ monolayer under the tensile strain of 2.5\% and 5\%. For the 2.5\% strain, both the formal $e'_g$$^2$ and $a_{1g}$$^1$$L_{z-}$$^1$ solutions are achieved in our calculations. The $e'_g$$^2$ solution has a spin moment of 1.87 $\mu_{\rm B}$ and an in-plane orbital moment of --0.44 $\mu_{\rm B}$, and however, it has a much higher total energy than the $a_{1g}$$^1$$L_{z-}$$^1$ solution by 80.9 meV/fu. Indeed, here the $a_{1g}$$^1$$L_{z-}$$^1$ solution becomes the ground state as expected, and it has a spin moment of 1.92 $\mu_{\rm B}$ and a large out-of-plane orbital moment of --1.16 $\mu_{\rm B}$, see Table II. This orbital moment tends via the SOC to align the spin moment along the $z$ axis, and if the spin moment flips into the $xy$ plane, there would be a large SOC energy cost. As a result, the $a_{1g}$$^1$$L_{z-}$$^1$ ground state has a large perpendicular magnetic anisotropy, which is up to 15.0 meV/fu by our LSDA+SOC+$U$ calculations. For the 5\% strain, the $a_{1g}$$^1$$L_{z-}$$^1$ ground state becomes even more stable against the formal $e'_g$$^2$ state by 141.6 meV/fu, and the resultant perpendicular magnetic anisotropy is 13.9 meV/fu. Apparently, the tensile strain can effectively modify the electronic ground state of VBr$_3$ monolayer and tune its magnetic anisotropy, from the formal $e'_g$$^2$ ground state and a moderate parallel magnetic anisotropy to the $a_{1g}$$^1$$L_{z-}$$^1$ ground state and a large perpendicular anisotropy.
 \begin{figure}[t]
\centering
\includegraphics[width=8.5cm]{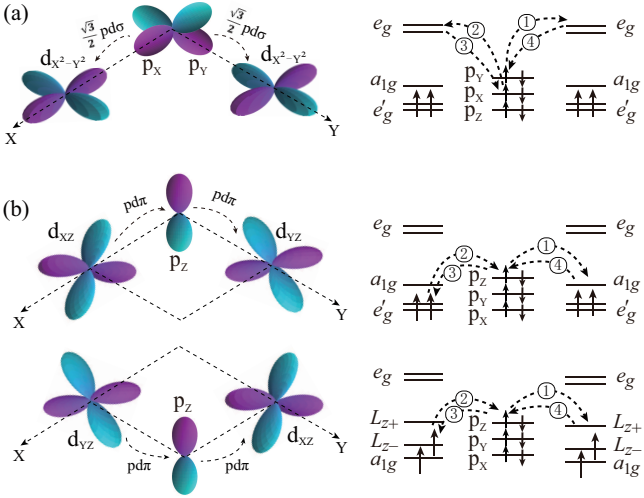}
 \caption{FM superexchange in the (near) $90^{\circ}$ V-Br-V bonds: (a) the charge transfer type and (b) the Mott-Hubbard type.
 }
\label{fig1:el}
\end{figure}

Moreover, the tensile strain turns out to enhance the FM coupling significantly: For the bare VBr$_3$ monolayer, the $e'_g$$^2$ ground state prefers the FM superexchange, which is more stable than the AF one by 7.9 meV/fu. For the monolayer under 2.5\% strain, the $a_{1g}$$^1$$L_{z-}$$^1$ ground state has a much enhanced FM stability against the AF by 31.8 meV/fu, which is up to 38.4 meV/fu for 5\% strain, see Table II. Now we provide a picture to understand why the FM superexchange is largely enhanced under the tensile strain, i.e., upon the electronic state transition from $e'_g$$^2$ to $a_{1g}$$^1$$L_{z-}$$^1$. To better discuss the electron hopping, here we use the local $XYZ$ coordinates of the VBr$_3$ octahedra. Then the V $3d$ wave functions can be written as follows
\begin{equation*}
\begin{aligned}
&\left|e_{g}\right\rangle=\frac{1}{\sqrt{2}}(\left|Z^{2}\right\rangle \pm \left| X^{2}-Y^{2}\right\rangle)\\
&\left|a_{1g}\right\rangle=\frac{1}{\sqrt{3}}(\left|XY\right\rangle
+\left|XZ\right\rangle+\left|YZ\right\rangle) \quad\quad\quad\quad\quad\quad(2)\\
&\left|e_{g}^{\prime}\right\rangle=L_{z\pm}=\frac{1}{\sqrt{3}}(\left|XY\right\rangle
+e^{\pm \frac{i2 \pi}{3}}\left|XZ\right\rangle+e^{\pm  \frac{i4 \pi}{3}}\left|YZ\right\rangle).
\end{aligned}
\end{equation*}
Having the early transition metal V and the strong V $3d$-Br $4p$ covalency, VBr$_3$ could be at the border of Mott-Hubbard type and charge-transfer one.~\cite{Khomskii_2001} When considering the charge-transfer behavior, the V $3d_{X^2-Y^2}-$Br $4p_{X,Y}-$V $3d_{X^2-Y^2}$ virtual hoppings (forth and back, large $pd\sigma$ type) with double holes on the ligand Br $4p$ orbitals are important, see Fig. 7(a), and this would give a FM superexchange which sees no difference between the $e'_g$$^2$ and $a_{1g}$$^1$$L_{z-}$$^1$ states. However, in the Mott-Hubbard regime, the V$^{3+}$ ions in the $e'_g$$^2$ or $a_{1g}$$^1$$L_{z-}$$^1$ state would undergo much different FM superexchange. Considering an effective $t_{2g}$ electron hopping in two neighboring V$^{3+}$ ions via the common ligand Br $4p$ orbital [see Fig. 7(b)], with the strength $t_{pd\pi}^2/\Delta$ where $\Delta$ is the charge transfer energy, the effective hopping between the $a_{1g}$ orbitals of two neighboring V$^{3+}$ ions is parameterized as 2$t_{pd\pi}^2/3\Delta$ (defined as 2$t_0$), and so is the hopping in between the $L_{z+}$ and $L_{z-}$ orbitals, and the rest inter-site hopping within the $t_{2g}$ is $t_0$=$t_{pd\pi}^2/3\Delta$, see more details in Refs. 21 and 33. Then, the superexchange FM stability against AF via the occupied to unoccupied $t_{2g}$ is (10$t_0^2/U$)(2$J_{\rm H}/U$) in the $a_{1g}$$^1$$L_{z-}$$^1$ state, being more than doubled as compared with that of (4$t_0^2/U$)(2$J_{\rm H}/U$) in the $e'_g$$^2$ state.

As shown above, the tensile strain significantly enhances the FM superexchange of the VBr$_3$ monolayer and the SIA. Therefore, the FM ordering temperature $T_{\rm C}$ of the VBr$_3$ monolayer would be largely increased by the tensile strain. We carry out Monte Carlo simulations to estimate the $T_{\rm C}$, using the following spin Hamiltonian
\begin{equation*}
H= -D \sum_{i}(S^z_i)^{2}-\frac{J}{2} \sum_{\langle i j\rangle} \overrightarrow{S_{i}}\cdot \overrightarrow{S_{j}}-\frac{J^{\prime}}{2} \sum_{\langle i j\rangle} S_{i}^{z} \cdot S_{j}^{z}, \quad(3)
\end{equation*}
where the first term stands for the SIA [positive (negative) $D$ for the easy (hard) $z$-axis magnetization], the second term describes the Heisenberg isotropic exchange [positive (negative) $J$ for FM (AF) coupling], and the last term represents the anisotropic exchange [positive (negative) $J^{\prime}$ for the easy (hard) $z$-axis magnetization]. The sum over $i$ runs over all V$^{3+}$ sites with $S$=1 in the honeycomb lattice, and $j$ over the three nearest neighbors of each $i$. The parameters $D$, $J$, and $J^{\prime}$ are determined by calculating the four magnetic states, i.e., the FM and AF states with the in-plane or out-of-plane magnetization, see Table II.

 \begin{figure}[t]
 \centering
\includegraphics[scale=0.8]{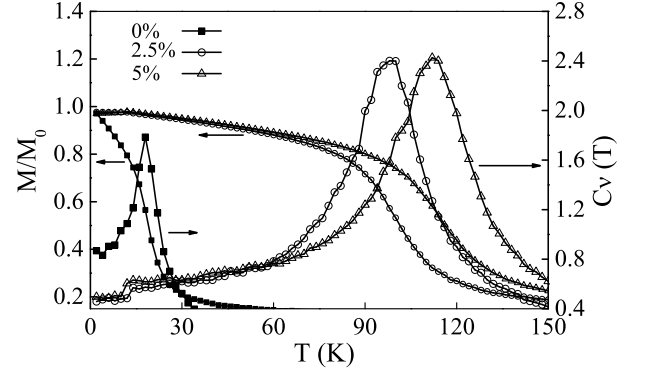}
 \caption{ Monte Carlo simulations of the magnetization and specific heat of the VBr$_{3}$ monolayer under different tensile strains.
 }
\label{fig1:el}
\end{figure}

For the VBr$_{3}$ monolayer, we find that $J^{\prime}$ is one to two orders of magnitude smaller than $D$ and $J$, showing a minor contribution  from the anisotropic exchange. Thus, the 2D FM order of VBr$_{3}$ monolayer is mainly stabilized by the FM superexchange and SIA. (This is different from the well studied CrI$_3$ monolayer, where the Cr$^{3+}$ ion has a closed $t_{2g}$$^3$ shell and has a negligible SIA, and the notable anisotropic exchange comes from the strong SOC of the heavy I $5p$ orbitals and strong Cr $3d$-I $5p$ hybridization.~\cite{Lado_2017,Kim_2019_PRL}) Note that for VBr$_{3}$ monolayer under the tensile strain, the FM $J$ parameter is significantly enhanced, and the SIA $D$ parameter changes from a small negative value to a large positive one, see Table II. For the bare VBr$_{3}$ monolayer, its $T_{\rm C}$ is about 20 K (see Fig. 8), according to our Monte Carlo simulations, due to the relatively small SIA and the relatively weak FM superexchange. In strong contrast, the VBr$_{3}$ monolayer has a significantly increasing $T_{\rm C}$ up to 100-115 K under the 2.5-5\% tensile strain, due to the strong FM coupling and strong SIA.
Therefore, we propose that the VBr$_{3}$ monolayer under a tensile strain would be a promising 2D FM semiconductor, which calls for an experimental verification.

\section{Summary}

In summary, using density functional calculations including SOC effect and electron correlation, and the crystal field and superexchange pictures as well, we find that the vdW material VBr$_{3}$ is in the formal $e'_{g}$$^2$ ground state and has an intralayer FM coupling and a weak parallel magnetic anisotropy. These results well account for the experimental magnetic behavior of the bulk VBr$_{3}$. More interestingly, the VBr$_{3}$ monolayer has a well comparable or even smaller cleavage energy as compared with other known 2D materials, and it is proposed to be a 2D FM semiconductor. We find that the bare VBr$_3$ monolayer has $T_{\rm C}$$\approx$20 K estimated from our Monte Carlo simulations, which is due to the relatively weak FM coupling and the small parallel magnetic anisotropy. In strong contrast, a tensile strain turns the VBr$_3$ monolayer into the $a_{1g}$$^1$$L_{z-}$$^1$ ground state. Then, the VBr$_3$ monolayer under the 2.5-5\% tensile strain has a largely increasing $T_{\rm C}$ up to 100-115 K, which arises from the significantly enhanced FM superexchange and the strong perpendicular magnetic anisotropy. Thus we conclude that VBr$_{3}$ monolayer under a tensile strain has a largely tunable magnetic anisotropy and quite high $T_{\rm C}$ and it would be a promising 2D FM semiconductor.

\section*{Acknowledgements}
This work was supported by the NSF of China (Grant No. 11674064) and by the National Key Research and Development Program of China (Grant No. 2016YFA0300700).

%This work was supported by the NSF of China (Grant No. 11674064) and by the National Key Research and Development Program of China (Grant No. 2016YFA0300700).

%%%END OF MAIN TEXT%%%

%The \balance command can be used to balance the columns on the final page if desired. It should be placed anywhere within the first column of the last page.

\nocite{xyz}

%If notes are included in your references you can change the title from 'References' to 'Notes and references' using the following command:
%\renewcommand\refname{References}

%%%REFERENCES%%%
\bibliography{rsc} %You need to replace "rsc" on this line with the name of your .bib file

%merlin.mbs apsrev4-1.bst 2010-07-25 4.21a (PWD, AO, DPC) hacked
%Control: key (0)
%Control: author (8) initials jnrlst
%Control: editor formatted (1) identically to author
%Control: production of article title (-1) disabled
%Control: page (0) single
%Control: year (1) truncated
%Control: production of eprint (0) enabled
\begin{thebibliography}{33}%
\makeatletter
\providecommand \@ifxundefined [1]{%
 \@ifx{#1\undefined}
}%
\providecommand \@ifnum [1]{%
 \ifnum #1\expandafter \@firstoftwo
 \else \expandafter \@secondoftwo
 \fi
}%
\providecommand \@ifx [1]{%
 \ifx #1\expandafter \@firstoftwo
 \else \expandafter \@secondoftwo
 \fi
}%
\providecommand \natexlab [1]{#1}%
\providecommand \enquote  [1]{``#1''}%
\providecommand \bibnamefont  [1]{#1}%
\providecommand \bibfnamefont [1]{#1}%
\providecommand \citenamefont [1]{#1}%
\providecommand \href@noop [0]{\@secondoftwo}%
\providecommand \href [0]{\begingroup \@sanitize@url \@href}%
\providecommand \@href[1]{\@@startlink{#1}\@@href}%
\providecommand \@@href[1]{\endgroup#1\@@endlink}%
\providecommand \@sanitize@url [0]{\catcode `\\12\catcode `\$12\catcode
  `\&12\catcode `\#12\catcode `\^12\catcode `\_12\catcode `\%12\relax}%
\providecommand \@@startlink[1]{}%
\providecommand \@@endlink[0]{}%
\providecommand \url  [0]{\begingroup\@sanitize@url \@url }%
\providecommand \@url [1]{\endgroup\@href {#1}{\urlprefix }}%
\providecommand \urlprefix  [0]{URL }%
\providecommand \Eprint [0]{\href }%
\providecommand \doibase [0]{http://dx.doi.org/}%
\providecommand \selectlanguage [0]{\@gobble}%
\providecommand \bibinfo  [0]{\@secondoftwo}%
\providecommand \bibfield  [0]{\@secondoftwo}%
\providecommand \translation [1]{[#1]}%
\providecommand \BibitemOpen [0]{}%
\providecommand \bibitemStop [0]{}%
\providecommand \bibitemNoStop [0]{.\EOS\space}%
\providecommand \EOS [0]{\spacefactor3000\relax}%
\providecommand \BibitemShut  [1]{\csname bibitem#1\endcsname}%
\let\auto@bib@innerbib\@empty
%</preamble>
\bibitem [{\citenamefont {Huang}\ \emph {et~al.}(2017)\citenamefont {Huang},
  \citenamefont {Clark}, \citenamefont {Navarro-Moratalla}, \citenamefont
  {Klein}, \citenamefont {Cheng}, \citenamefont {Seyler}, \citenamefont
  {Zhong}, \citenamefont {Schmidgall}, \citenamefont {McGuire}, \citenamefont
  {Cobden}, \citenamefont {Yao}, \citenamefont {Xiao}, \citenamefont
  {Jarillo-Herrero},\ and\ \citenamefont {Xu}}]{Huang_2017}%
  \BibitemOpen
  \bibfield  {author} {\bibinfo {author} {\bibfnamefont {B.}~\bibnamefont
  {Huang}}, \bibinfo {author} {\bibfnamefont {G.}~\bibnamefont {Clark}},
  \bibinfo {author} {\bibfnamefont {E.}~\bibnamefont {Navarro-Moratalla}},
  \bibinfo {author} {\bibfnamefont {D.~R.}\ \bibnamefont {Klein}}, \bibinfo
  {author} {\bibfnamefont {R.}~\bibnamefont {Cheng}}, \bibinfo {author}
  {\bibfnamefont {K.~L.}\ \bibnamefont {Seyler}}, \bibinfo {author}
  {\bibfnamefont {D.}~\bibnamefont {Zhong}}, \bibinfo {author} {\bibfnamefont
  {E.}~\bibnamefont {Schmidgall}}, \bibinfo {author} {\bibfnamefont {M.~A.}\
  \bibnamefont {McGuire}}, \bibinfo {author} {\bibfnamefont {D.~H.}\
  \bibnamefont {Cobden}}, \bibinfo {author} {\bibfnamefont {W.}~\bibnamefont
  {Yao}}, \bibinfo {author} {\bibfnamefont {D.}~\bibnamefont {Xiao}}, \bibinfo
  {author} {\bibfnamefont {P.}~\bibnamefont {Jarillo-Herrero}}, \ and\ \bibinfo
  {author} {\bibfnamefont {X.}~\bibnamefont {Xu}},\ }\href {\doibase
  10.1038/nature22391} {\bibfield  {journal} {\bibinfo  {journal} {Nature}\
  }\textbf {\bibinfo {volume} {546}},\ \bibinfo {pages} {270} (\bibinfo {year}
  {2017})}\BibitemShut {NoStop}%
\bibitem [{\citenamefont {Gong}\ \emph {et~al.}(2017)\citenamefont {Gong},
  \citenamefont {Li}, \citenamefont {Li}, \citenamefont {Ji}, \citenamefont
  {Stern}, \citenamefont {Xia}, \citenamefont {Cao}, \citenamefont {Bao},
  \citenamefont {Wang}, \citenamefont {Wang}, \citenamefont {Qiu},
  \citenamefont {Cava}, \citenamefont {Louie}, \citenamefont {Xia},\ and\
  \citenamefont {Zhang}}]{Gong_2017}%
  \BibitemOpen
  \bibfield  {author} {\bibinfo {author} {\bibfnamefont {C.}~\bibnamefont
  {Gong}}, \bibinfo {author} {\bibfnamefont {L.}~\bibnamefont {Li}}, \bibinfo
  {author} {\bibfnamefont {Z.}~\bibnamefont {Li}}, \bibinfo {author}
  {\bibfnamefont {H.}~\bibnamefont {Ji}}, \bibinfo {author} {\bibfnamefont
  {A.}~\bibnamefont {Stern}}, \bibinfo {author} {\bibfnamefont
  {Y.}~\bibnamefont {Xia}}, \bibinfo {author} {\bibfnamefont {T.}~\bibnamefont
  {Cao}}, \bibinfo {author} {\bibfnamefont {W.}~\bibnamefont {Bao}}, \bibinfo
  {author} {\bibfnamefont {C.}~\bibnamefont {Wang}}, \bibinfo {author}
  {\bibfnamefont {Y.}~\bibnamefont {Wang}}, \bibinfo {author} {\bibfnamefont
  {Z.~Q.}\ \bibnamefont {Qiu}}, \bibinfo {author} {\bibfnamefont {R.~J.}\
  \bibnamefont {Cava}}, \bibinfo {author} {\bibfnamefont {S.~G.}\ \bibnamefont
  {Louie}}, \bibinfo {author} {\bibfnamefont {J.}~\bibnamefont {Xia}}, \ and\
  \bibinfo {author} {\bibfnamefont {X.}~\bibnamefont {Zhang}},\ }\href
  {\doibase 10.1038/nature22060} {\bibfield  {journal} {\bibinfo  {journal}
  {Nature}\ }\textbf {\bibinfo {volume} {546}},\ \bibinfo {pages} {265}
  (\bibinfo {year} {2017})}\BibitemShut {NoStop}%
\bibitem [{\citenamefont {Deng}\ \emph {et~al.}(2018)\citenamefont {Deng},
  \citenamefont {Yu}, \citenamefont {Song}, \citenamefont {Zhang},
  \citenamefont {Wang}, \citenamefont {Sun}, \citenamefont {Yi}, \citenamefont
  {Wu}, \citenamefont {Wu}, \citenamefont {Zhu}, \citenamefont {Wang},
  \citenamefont {Chen},\ and\ \citenamefont {Zhang}}]{Deng_2018}%
  \BibitemOpen
  \bibfield  {author} {\bibinfo {author} {\bibfnamefont {Y.}~\bibnamefont
  {Deng}}, \bibinfo {author} {\bibfnamefont {Y.}~\bibnamefont {Yu}}, \bibinfo
  {author} {\bibfnamefont {Y.}~\bibnamefont {Song}}, \bibinfo {author}
  {\bibfnamefont {J.}~\bibnamefont {Zhang}}, \bibinfo {author} {\bibfnamefont
  {N.~Z.}\ \bibnamefont {Wang}}, \bibinfo {author} {\bibfnamefont
  {Z.}~\bibnamefont {Sun}}, \bibinfo {author} {\bibfnamefont {Y.}~\bibnamefont
  {Yi}}, \bibinfo {author} {\bibfnamefont {Y.~Z.}\ \bibnamefont {Wu}}, \bibinfo
  {author} {\bibfnamefont {S.}~\bibnamefont {Wu}}, \bibinfo {author}
  {\bibfnamefont {J.}~\bibnamefont {Zhu}}, \bibinfo {author} {\bibfnamefont
  {J.}~\bibnamefont {Wang}}, \bibinfo {author} {\bibfnamefont {X.~H.}\
  \bibnamefont {Chen}}, \ and\ \bibinfo {author} {\bibfnamefont
  {Y.}~\bibnamefont {Zhang}},\ }\href {\doibase 10.1038/s41586-018-0626-9}
  {\bibfield  {journal} {\bibinfo  {journal} {Nature}\ }\textbf {\bibinfo
  {volume} {563}},\ \bibinfo {pages} {94} (\bibinfo {year} {2018})}\BibitemShut
  {NoStop}%
\bibitem [{\citenamefont {Chen}\ \emph {et~al.}(2019)\citenamefont {Chen},
  \citenamefont {Sun}, \citenamefont {Wang}, \citenamefont {Gu}, \citenamefont
  {Xu}, \citenamefont {Wu},\ and\ \citenamefont {Gao}}]{Chen_2019}%
  \BibitemOpen
  \bibfield  {author} {\bibinfo {author} {\bibfnamefont {W.}~\bibnamefont
  {Chen}}, \bibinfo {author} {\bibfnamefont {Z.}~\bibnamefont {Sun}}, \bibinfo
  {author} {\bibfnamefont {Z.}~\bibnamefont {Wang}}, \bibinfo {author}
  {\bibfnamefont {L.}~\bibnamefont {Gu}}, \bibinfo {author} {\bibfnamefont
  {X.}~\bibnamefont {Xu}}, \bibinfo {author} {\bibfnamefont {S.}~\bibnamefont
  {Wu}}, \ and\ \bibinfo {author} {\bibfnamefont {C.}~\bibnamefont {Gao}},\
  }\href {\doibase 10.1126/science.aav1937} {\bibfield  {journal} {\bibinfo
  {journal} {Science}\ }\textbf {\bibinfo {volume} {366}},\ \bibinfo {pages}
  {983} (\bibinfo {year} {2019})}\BibitemShut {NoStop}%
\bibitem [{\citenamefont {Li}\ \emph {et~al.}(2019)\citenamefont {Li},
  \citenamefont {Jiang}, \citenamefont {Sivadas}, \citenamefont {Wang},
  \citenamefont {Xu}, \citenamefont {Weber}, \citenamefont {Goldberger},
  \citenamefont {Watanabe}, \citenamefont {Taniguchi}, \citenamefont {Fennie},
  \citenamefont {Mak},\ and\ \citenamefont {Shan}}]{Li_2019}%
  \BibitemOpen
  \bibfield  {author} {\bibinfo {author} {\bibfnamefont {T.}~\bibnamefont
  {Li}}, \bibinfo {author} {\bibfnamefont {S.}~\bibnamefont {Jiang}}, \bibinfo
  {author} {\bibfnamefont {N.}~\bibnamefont {Sivadas}}, \bibinfo {author}
  {\bibfnamefont {Z.}~\bibnamefont {Wang}}, \bibinfo {author} {\bibfnamefont
  {Y.}~\bibnamefont {Xu}}, \bibinfo {author} {\bibfnamefont {D.}~\bibnamefont
  {Weber}}, \bibinfo {author} {\bibfnamefont {J.~E.}\ \bibnamefont
  {Goldberger}}, \bibinfo {author} {\bibfnamefont {K.}~\bibnamefont
  {Watanabe}}, \bibinfo {author} {\bibfnamefont {T.}~\bibnamefont {Taniguchi}},
  \bibinfo {author} {\bibfnamefont {C.~J.}\ \bibnamefont {Fennie}}, \bibinfo
  {author} {\bibfnamefont {K.~F.}\ \bibnamefont {Mak}}, \ and\ \bibinfo
  {author} {\bibfnamefont {J.}~\bibnamefont {Shan}},\ }\href {\doibase
  10.1038/s41563-019-0506-1} {\bibfield  {journal} {\bibinfo  {journal} {Nat.
  Mater.}\ }\textbf {\bibinfo {volume} {18}},\ \bibinfo {pages} {1303}
  (\bibinfo {year} {2019})}\BibitemShut {NoStop}%
\bibitem [{\citenamefont {Song}\ \emph {et~al.}(2019)\citenamefont {Song},
  \citenamefont {Fei}, \citenamefont {Yankowitz}, \citenamefont {Lin},
  \citenamefont {Jiang}, \citenamefont {Hwangbo}, \citenamefont {Zhang},
  \citenamefont {Sun}, \citenamefont {Taniguchi}, \citenamefont {Watanabe},
  \citenamefont {McGuire}, \citenamefont {Graf}, \citenamefont {Cao},
  \citenamefont {Chu}, \citenamefont {Cobden}, \citenamefont {Dean},
  \citenamefont {Xiao},\ and\ \citenamefont {Xu}}]{Song_2019}%
  \BibitemOpen
  \bibfield  {author} {\bibinfo {author} {\bibfnamefont {T.}~\bibnamefont
  {Song}}, \bibinfo {author} {\bibfnamefont {Z.}~\bibnamefont {Fei}}, \bibinfo
  {author} {\bibfnamefont {M.}~\bibnamefont {Yankowitz}}, \bibinfo {author}
  {\bibfnamefont {Z.}~\bibnamefont {Lin}}, \bibinfo {author} {\bibfnamefont
  {Q.}~\bibnamefont {Jiang}}, \bibinfo {author} {\bibfnamefont
  {K.}~\bibnamefont {Hwangbo}}, \bibinfo {author} {\bibfnamefont
  {Q.}~\bibnamefont {Zhang}}, \bibinfo {author} {\bibfnamefont
  {B.}~\bibnamefont {Sun}}, \bibinfo {author} {\bibfnamefont {T.}~\bibnamefont
  {Taniguchi}}, \bibinfo {author} {\bibfnamefont {K.}~\bibnamefont {Watanabe}},
  \bibinfo {author} {\bibfnamefont {M.~A.}\ \bibnamefont {McGuire}}, \bibinfo
  {author} {\bibfnamefont {D.}~\bibnamefont {Graf}}, \bibinfo {author}
  {\bibfnamefont {T.}~\bibnamefont {Cao}}, \bibinfo {author} {\bibfnamefont
  {J.-H.}\ \bibnamefont {Chu}}, \bibinfo {author} {\bibfnamefont {D.~H.}\
  \bibnamefont {Cobden}}, \bibinfo {author} {\bibfnamefont {C.~R.}\
  \bibnamefont {Dean}}, \bibinfo {author} {\bibfnamefont {D.}~\bibnamefont
  {Xiao}}, \ and\ \bibinfo {author} {\bibfnamefont {X.}~\bibnamefont {Xu}},\
  }\href {\doibase 10.1038/s41563-019-0505-2} {\bibfield  {journal} {\bibinfo
  {journal} {Nat. Mater.}\ }\textbf {\bibinfo {volume} {18}},\ \bibinfo {pages}
  {1298} (\bibinfo {year} {2019})}\BibitemShut {NoStop}%
\bibitem [{\citenamefont {Klein}\ \emph {et~al.}(2019)\citenamefont {Klein},
  \citenamefont {MacNeill}, \citenamefont {Song}, \citenamefont {Larson},
  \citenamefont {Fang}, \citenamefont {Xu}, \citenamefont {Ribeiro},
  \citenamefont {Canfield}, \citenamefont {Kaxiras}, \citenamefont {Comin},\
  and\ \citenamefont {Jarillo-Herrero}}]{Klein_2019}%
  \BibitemOpen
  \bibfield  {author} {\bibinfo {author} {\bibfnamefont {D.~R.}\ \bibnamefont
  {Klein}}, \bibinfo {author} {\bibfnamefont {D.}~\bibnamefont {MacNeill}},
  \bibinfo {author} {\bibfnamefont {Q.}~\bibnamefont {Song}}, \bibinfo {author}
  {\bibfnamefont {D.~T.}\ \bibnamefont {Larson}}, \bibinfo {author}
  {\bibfnamefont {S.}~\bibnamefont {Fang}}, \bibinfo {author} {\bibfnamefont
  {M.}~\bibnamefont {Xu}}, \bibinfo {author} {\bibfnamefont {R.~A.}\
  \bibnamefont {Ribeiro}}, \bibinfo {author} {\bibfnamefont {P.~C.}\
  \bibnamefont {Canfield}}, \bibinfo {author} {\bibfnamefont {E.}~\bibnamefont
  {Kaxiras}}, \bibinfo {author} {\bibfnamefont {R.}~\bibnamefont {Comin}}, \
  and\ \bibinfo {author} {\bibfnamefont {P.}~\bibnamefont {Jarillo-Herrero}},\
  }\href {\doibase 10.1038/s41567-019-0651-0} {\bibfield  {journal} {\bibinfo
  {journal} {Nat. Phys.}\ }\textbf {\bibinfo {volume} {15}},\ \bibinfo {pages}
  {1255} (\bibinfo {year} {2019})}\BibitemShut {NoStop}%
\bibitem [{\citenamefont {Huang}\ \emph {et~al.}(2020)\citenamefont {Huang},
  \citenamefont {Cenker}, \citenamefont {Zhang}, \citenamefont {Ray},
  \citenamefont {Song}, \citenamefont {Taniguchi}, \citenamefont {Watanabe},
  \citenamefont {McGuire}, \citenamefont {Xiao},\ and\ \citenamefont
  {Xu}}]{Huang_2020}%
  \BibitemOpen
  \bibfield  {author} {\bibinfo {author} {\bibfnamefont {B.}~\bibnamefont
  {Huang}}, \bibinfo {author} {\bibfnamefont {J.}~\bibnamefont {Cenker}},
  \bibinfo {author} {\bibfnamefont {X.}~\bibnamefont {Zhang}}, \bibinfo
  {author} {\bibfnamefont {E.~L.}\ \bibnamefont {Ray}}, \bibinfo {author}
  {\bibfnamefont {T.}~\bibnamefont {Song}}, \bibinfo {author} {\bibfnamefont
  {T.}~\bibnamefont {Taniguchi}}, \bibinfo {author} {\bibfnamefont
  {K.}~\bibnamefont {Watanabe}}, \bibinfo {author} {\bibfnamefont {M.~A.}\
  \bibnamefont {McGuire}}, \bibinfo {author} {\bibfnamefont {D.}~\bibnamefont
  {Xiao}}, \ and\ \bibinfo {author} {\bibfnamefont {X.}~\bibnamefont {Xu}},\
  }\href {\doibase 10.1038/s41565-019-0598-4} {\bibfield  {journal} {\bibinfo
  {journal} {Nat. Nanotechnol.}\ }\textbf {\bibinfo {volume} {15}},\ \bibinfo
  {pages} {212} (\bibinfo {year} {2020})}\BibitemShut {NoStop}%
\bibitem [{\citenamefont {Kim}\ \emph {et~al.}(2019{\natexlab{a}})\citenamefont
  {Kim}, \citenamefont {Yang}, \citenamefont {Li}, \citenamefont {Jiang},
  \citenamefont {Jin}, \citenamefont {Tao}, \citenamefont {Nichols},
  \citenamefont {Sfigakis}, \citenamefont {Zhong}, \citenamefont {Li},
  \citenamefont {Tian}, \citenamefont {Cory}, \citenamefont {Miao},
  \citenamefont {Shan}, \citenamefont {Mak}, \citenamefont {Lei}, \citenamefont
  {Sun}, \citenamefont {Zhao},\ and\ \citenamefont {Tsen}}]{Kim_2019}%
  \BibitemOpen
  \bibfield  {author} {\bibinfo {author} {\bibfnamefont {H.~H.}\ \bibnamefont
  {Kim}}, \bibinfo {author} {\bibfnamefont {B.}~\bibnamefont {Yang}}, \bibinfo
  {author} {\bibfnamefont {S.}~\bibnamefont {Li}}, \bibinfo {author}
  {\bibfnamefont {S.}~\bibnamefont {Jiang}}, \bibinfo {author} {\bibfnamefont
  {C.}~\bibnamefont {Jin}}, \bibinfo {author} {\bibfnamefont {Z.}~\bibnamefont
  {Tao}}, \bibinfo {author} {\bibfnamefont {G.}~\bibnamefont {Nichols}},
  \bibinfo {author} {\bibfnamefont {F.}~\bibnamefont {Sfigakis}}, \bibinfo
  {author} {\bibfnamefont {S.}~\bibnamefont {Zhong}}, \bibinfo {author}
  {\bibfnamefont {C.}~\bibnamefont {Li}}, \bibinfo {author} {\bibfnamefont
  {S.}~\bibnamefont {Tian}}, \bibinfo {author} {\bibfnamefont {D.~G.}\
  \bibnamefont {Cory}}, \bibinfo {author} {\bibfnamefont {G.-X.}\ \bibnamefont
  {Miao}}, \bibinfo {author} {\bibfnamefont {J.}~\bibnamefont {Shan}}, \bibinfo
  {author} {\bibfnamefont {K.~F.}\ \bibnamefont {Mak}}, \bibinfo {author}
  {\bibfnamefont {H.}~\bibnamefont {Lei}}, \bibinfo {author} {\bibfnamefont
  {K.}~\bibnamefont {Sun}}, \bibinfo {author} {\bibfnamefont {L.}~\bibnamefont
  {Zhao}}, \ and\ \bibinfo {author} {\bibfnamefont {A.~W.}\ \bibnamefont
  {Tsen}},\ }\href {\doibase 10.1073/pnas.1902100116} {\bibfield  {journal}
  {\bibinfo  {journal} {Proc. Natl. Acad. Sci. U. S. A.}\ }\textbf {\bibinfo
  {volume} {116}},\ \bibinfo {pages} {11131} (\bibinfo {year}
  {2019}{\natexlab{a}})},\ \Eprint
  {http://arxiv.org/abs/https://www.pnas.org/content/116/23/11131.full.pdf}
  {https://www.pnas.org/content/116/23/11131.full.pdf} \BibitemShut {NoStop}%
\bibitem [{\citenamefont {Mermin}\ and\ \citenamefont {Wagner}(1966)}]{MW}%
  \BibitemOpen
  \bibfield  {author} {\bibinfo {author} {\bibfnamefont {N.~D.}\ \bibnamefont
  {Mermin}}\ and\ \bibinfo {author} {\bibfnamefont {H.}~\bibnamefont
  {Wagner}},\ }\href {\doibase 10.1103/PhysRevLett.17.1133} {\bibfield
  {journal} {\bibinfo  {journal} {Phys. Rev. Lett.}\ }\textbf {\bibinfo
  {volume} {17}},\ \bibinfo {pages} {1133} (\bibinfo {year}
  {1966})}\BibitemShut {NoStop}%
\bibitem [{\citenamefont {Cardoso}\ \emph {et~al.}(2018)\citenamefont
  {Cardoso}, \citenamefont {Soriano}, \citenamefont {Garcia-Martinez},\ and\
  \citenamefont {Fernandez-Rossier}}]{valves}%
  \BibitemOpen
  \bibfield  {author} {\bibinfo {author} {\bibfnamefont {C.}~\bibnamefont
  {Cardoso}}, \bibinfo {author} {\bibfnamefont {D.}~\bibnamefont {Soriano}},
  \bibinfo {author} {\bibfnamefont {N.~A.}\ \bibnamefont {Garcia-Martinez}}, \
  and\ \bibinfo {author} {\bibfnamefont {J.}~\bibnamefont
  {Fernandez-Rossier}},\ }\href {\doibase 10.1103/PhysRevLett.121.067701}
  {\bibfield  {journal} {\bibinfo  {journal} {Phys. Rev. Lett.}\ }\textbf
  {\bibinfo {volume} {121}},\ \bibinfo {pages} {067701} (\bibinfo {year}
  {2018})}\BibitemShut {NoStop}%
\bibitem [{\citenamefont {Klein}\ \emph {et~al.}(2018)\citenamefont {Klein},
  \citenamefont {MacNeill}, \citenamefont {Lado}, \citenamefont {Soriano},
  \citenamefont {Navarro-Moratalla}, \citenamefont {Watanabe}, \citenamefont
  {Taniguchi}, \citenamefont {Manni}, \citenamefont {Canfield}, \citenamefont
  {Fernandez-Rossier},\ and\ \citenamefont {Jarillo-Herrero}}]{filter1}%
  \BibitemOpen
  \bibfield  {author} {\bibinfo {author} {\bibfnamefont {D.~R.}\ \bibnamefont
  {Klein}}, \bibinfo {author} {\bibfnamefont {D.}~\bibnamefont {MacNeill}},
  \bibinfo {author} {\bibfnamefont {J.~L.}\ \bibnamefont {Lado}}, \bibinfo
  {author} {\bibfnamefont {D.}~\bibnamefont {Soriano}}, \bibinfo {author}
  {\bibfnamefont {E.}~\bibnamefont {Navarro-Moratalla}}, \bibinfo {author}
  {\bibfnamefont {K.}~\bibnamefont {Watanabe}}, \bibinfo {author}
  {\bibfnamefont {T.}~\bibnamefont {Taniguchi}}, \bibinfo {author}
  {\bibfnamefont {S.}~\bibnamefont {Manni}}, \bibinfo {author} {\bibfnamefont
  {P.}~\bibnamefont {Canfield}}, \bibinfo {author} {\bibfnamefont
  {J.}~\bibnamefont {Fernandez-Rossier}}, \ and\ \bibinfo {author}
  {\bibfnamefont {P.}~\bibnamefont {Jarillo-Herrero}},\ }\href {\doibase
  10.1126/science.aar3617} {\bibfield  {journal} {\bibinfo  {journal}
  {Science}\ }\textbf {\bibinfo {volume} {360}},\ \bibinfo {pages} {1218}
  (\bibinfo {year} {2018})}\BibitemShut {NoStop}%
\bibitem [{\citenamefont {Song}\ \emph {et~al.}(2018)\citenamefont {Song},
  \citenamefont {Cai}, \citenamefont {Tu}, \citenamefont {Zhang}, \citenamefont
  {Huang}, \citenamefont {Wilson}, \citenamefont {Seyler}, \citenamefont {Zhu},
  \citenamefont {Taniguchi}, \citenamefont {Watanabe}, \citenamefont {McGuire},
  \citenamefont {Cobden}, \citenamefont {Xiao}, \citenamefont {Yao},\ and\
  \citenamefont {Xu}}]{filter2}%
  \BibitemOpen
  \bibfield  {author} {\bibinfo {author} {\bibfnamefont {T.}~\bibnamefont
  {Song}}, \bibinfo {author} {\bibfnamefont {X.}~\bibnamefont {Cai}}, \bibinfo
  {author} {\bibfnamefont {M.~W.-Y.}\ \bibnamefont {Tu}}, \bibinfo {author}
  {\bibfnamefont {X.}~\bibnamefont {Zhang}}, \bibinfo {author} {\bibfnamefont
  {B.}~\bibnamefont {Huang}}, \bibinfo {author} {\bibfnamefont {N.~P.}\
  \bibnamefont {Wilson}}, \bibinfo {author} {\bibfnamefont {K.~L.}\
  \bibnamefont {Seyler}}, \bibinfo {author} {\bibfnamefont {L.}~\bibnamefont
  {Zhu}}, \bibinfo {author} {\bibfnamefont {T.}~\bibnamefont {Taniguchi}},
  \bibinfo {author} {\bibfnamefont {K.}~\bibnamefont {Watanabe}}, \bibinfo
  {author} {\bibfnamefont {M.~A.}\ \bibnamefont {McGuire}}, \bibinfo {author}
  {\bibfnamefont {D.~H.}\ \bibnamefont {Cobden}}, \bibinfo {author}
  {\bibfnamefont {D.}~\bibnamefont {Xiao}}, \bibinfo {author} {\bibfnamefont
  {W.}~\bibnamefont {Yao}}, \ and\ \bibinfo {author} {\bibfnamefont
  {X.}~\bibnamefont {Xu}},\ }\href {\doibase 10.1126/science.aar4851}
  {\bibfield  {journal} {\bibinfo  {journal} {Science}\ }\textbf {\bibinfo
  {volume} {360}},\ \bibinfo {pages} {1214} (\bibinfo {year}
  {2018})}\BibitemShut {NoStop}%
\bibitem [{\citenamefont {Soumyanarayanan}\ \emph {et~al.}(2016)\citenamefont
  {Soumyanarayanan}, \citenamefont {Reyren}, \citenamefont {Fert},\ and\
  \citenamefont {Panagopoulos}}]{storage}%
  \BibitemOpen
  \bibfield  {author} {\bibinfo {author} {\bibfnamefont {A.}~\bibnamefont
  {Soumyanarayanan}}, \bibinfo {author} {\bibfnamefont {N.}~\bibnamefont
  {Reyren}}, \bibinfo {author} {\bibfnamefont {A.}~\bibnamefont {Fert}}, \ and\
  \bibinfo {author} {\bibfnamefont {C.}~\bibnamefont {Panagopoulos}},\ }\href
  {\doibase 10.1038/nature19820} {\bibfield  {journal} {\bibinfo  {journal}
  {Nature}\ }\textbf {\bibinfo {volume} {539}},\ \bibinfo {pages} {509}
  (\bibinfo {year} {2016})}\BibitemShut {NoStop}%
\bibitem [{\citenamefont {Kong}\ \emph
  {et~al.}(2019{\natexlab{a}})\citenamefont {Kong}, \citenamefont {Guo},
  \citenamefont {Ni},\ and\ \citenamefont {Cava}}]{Kong_VBr}%
  \BibitemOpen
  \bibfield  {author} {\bibinfo {author} {\bibfnamefont {T.}~\bibnamefont
  {Kong}}, \bibinfo {author} {\bibfnamefont {S.}~\bibnamefont {Guo}}, \bibinfo
  {author} {\bibfnamefont {D.}~\bibnamefont {Ni}}, \ and\ \bibinfo {author}
  {\bibfnamefont {R.~J.}\ \bibnamefont {Cava}},\ }\href {\doibase
  10.1103/PhysRevMaterials.3.084419} {\bibfield  {journal} {\bibinfo  {journal}
  {Phys. Rev. Mater.}\ }\textbf {\bibinfo {volume} {3}},\ \bibinfo {pages}
  {084419} (\bibinfo {year} {2019}{\natexlab{a}})}\BibitemShut {NoStop}%
\bibitem [{\citenamefont {Kong}\ \emph
  {et~al.}(2019{\natexlab{b}})\citenamefont {Kong}, \citenamefont {Stolze},
  \citenamefont {Timmons}, \citenamefont {Tao}, \citenamefont {Ni},
  \citenamefont {Guo}, \citenamefont {Yang}, \citenamefont {Prozorov},\ and\
  \citenamefont {Cava}}]{Kong_2019}%
  \BibitemOpen
  \bibfield  {author} {\bibinfo {author} {\bibfnamefont {T.}~\bibnamefont
  {Kong}}, \bibinfo {author} {\bibfnamefont {K.}~\bibnamefont {Stolze}},
  \bibinfo {author} {\bibfnamefont {E.~I.}\ \bibnamefont {Timmons}}, \bibinfo
  {author} {\bibfnamefont {J.}~\bibnamefont {Tao}}, \bibinfo {author}
  {\bibfnamefont {D.}~\bibnamefont {Ni}}, \bibinfo {author} {\bibfnamefont
  {S.}~\bibnamefont {Guo}}, \bibinfo {author} {\bibfnamefont {Z.}~\bibnamefont
  {Yang}}, \bibinfo {author} {\bibfnamefont {R.}~\bibnamefont {Prozorov}}, \
  and\ \bibinfo {author} {\bibfnamefont {R.~J.}\ \bibnamefont {Cava}},\ }\href
  {\doibase 10.1002/adma.201808074} {\bibfield  {journal} {\bibinfo  {journal}
  {Adv. Mater.}\ }\textbf {\bibinfo {volume} {31}},\ \bibinfo {pages} {1808074}
  (\bibinfo {year} {2019}{\natexlab{b}})},\ \Eprint
  {http://arxiv.org/abs/https://onlinelibrary.wiley.com/doi/pdf/10.1002/adma.201808074}
  {https://onlinelibrary.wiley.com/doi/pdf/10.1002/adma.201808074} \BibitemShut
  {NoStop}%
\bibitem [{\citenamefont {Tian}\ \emph {et~al.}(2019)\citenamefont {Tian},
  \citenamefont {Zhang}, \citenamefont {Li}, \citenamefont {Ying},
  \citenamefont {Li}, \citenamefont {Zhang}, \citenamefont {Liu},\ and\
  \citenamefont {Lei}}]{Tian_2019_jacs}%
  \BibitemOpen
  \bibfield  {author} {\bibinfo {author} {\bibfnamefont {S.}~\bibnamefont
  {Tian}}, \bibinfo {author} {\bibfnamefont {J.-F.}\ \bibnamefont {Zhang}},
  \bibinfo {author} {\bibfnamefont {C.}~\bibnamefont {Li}}, \bibinfo {author}
  {\bibfnamefont {T.}~\bibnamefont {Ying}}, \bibinfo {author} {\bibfnamefont
  {S.}~\bibnamefont {Li}}, \bibinfo {author} {\bibfnamefont {X.}~\bibnamefont
  {Zhang}}, \bibinfo {author} {\bibfnamefont {K.}~\bibnamefont {Liu}}, \ and\
  \bibinfo {author} {\bibfnamefont {H.}~\bibnamefont {Lei}},\ }\href {\doibase
  10.1021/jacs.8b13584} {\bibfield  {journal} {\bibinfo  {journal} {J. Am.
  Chem. Soc.}\ }\textbf {\bibinfo {volume} {141}},\ \bibinfo {pages} {5326}
  (\bibinfo {year} {2019})}\BibitemShut {NoStop}%
\bibitem [{\citenamefont {Son}\ \emph {et~al.}(2019)\citenamefont {Son},
  \citenamefont {Coak}, \citenamefont {Lee}, \citenamefont {Kim}, \citenamefont
  {Kim}, \citenamefont {Hamidov}, \citenamefont {Cho}, \citenamefont {Liu},
  \citenamefont {Jarvis}, \citenamefont {Brown}, \citenamefont {Kim},
  \citenamefont {Park}, \citenamefont {Khomskii}, \citenamefont {Saxena},\ and\
  \citenamefont {Park}}]{Son_2019}%
  \BibitemOpen
  \bibfield  {author} {\bibinfo {author} {\bibfnamefont {S.}~\bibnamefont
  {Son}}, \bibinfo {author} {\bibfnamefont {M.~J.}\ \bibnamefont {Coak}},
  \bibinfo {author} {\bibfnamefont {N.}~\bibnamefont {Lee}}, \bibinfo {author}
  {\bibfnamefont {J.}~\bibnamefont {Kim}}, \bibinfo {author} {\bibfnamefont
  {T.~Y.}\ \bibnamefont {Kim}}, \bibinfo {author} {\bibfnamefont
  {H.}~\bibnamefont {Hamidov}}, \bibinfo {author} {\bibfnamefont
  {H.}~\bibnamefont {Cho}}, \bibinfo {author} {\bibfnamefont {C.}~\bibnamefont
  {Liu}}, \bibinfo {author} {\bibfnamefont {D.~M.}\ \bibnamefont {Jarvis}},
  \bibinfo {author} {\bibfnamefont {P.~A.~C.}\ \bibnamefont {Brown}}, \bibinfo
  {author} {\bibfnamefont {J.~H.}\ \bibnamefont {Kim}}, \bibinfo {author}
  {\bibfnamefont {C.-H.}\ \bibnamefont {Park}}, \bibinfo {author}
  {\bibfnamefont {D.~I.}\ \bibnamefont {Khomskii}}, \bibinfo {author}
  {\bibfnamefont {S.~S.}\ \bibnamefont {Saxena}}, \ and\ \bibinfo {author}
  {\bibfnamefont {J.-G.}\ \bibnamefont {Park}},\ }\href {\doibase
  10.1103/PhysRevB.99.041402} {\bibfield  {journal} {\bibinfo  {journal} {Phys.
  Rev. B}\ }\textbf {\bibinfo {volume} {99}},\ \bibinfo {pages} {041402(R)}
  (\bibinfo {year} {2019})}\BibitemShut {NoStop}%
\bibitem [{\citenamefont {Lado}\ and\ \citenamefont
  {Fernandez-Rossier}(2017)}]{Lado_2017}%
  \BibitemOpen
  \bibfield  {author} {\bibinfo {author} {\bibfnamefont {J.~L.}\ \bibnamefont
  {Lado}}\ and\ \bibinfo {author} {\bibfnamefont {J.}~\bibnamefont
  {Fernandez-Rossier}},\ }\href {\doibase 10.1088/2053-1583/aa75ed} {\bibfield
  {journal} {\bibinfo  {journal} {2D Mater.}\ }\textbf {\bibinfo {volume}
  {4}},\ \bibinfo {pages} {035002} (\bibinfo {year} {2017})}\BibitemShut
  {NoStop}%
\bibitem [{\citenamefont {Kim}\ \emph {et~al.}(2019{\natexlab{b}})\citenamefont
  {Kim}, \citenamefont {Kim}, \citenamefont {Ko}, \citenamefont {Seo},
  \citenamefont {Kim}, \citenamefont {Jang}, \citenamefont {Kim}, \citenamefont
  {Kim}, \citenamefont {Cheong},\ and\ \citenamefont {Park}}]{Kim_2019_PRL}%
  \BibitemOpen
  \bibfield  {author} {\bibinfo {author} {\bibfnamefont {D.-H.}\ \bibnamefont
  {Kim}}, \bibinfo {author} {\bibfnamefont {K.}~\bibnamefont {Kim}}, \bibinfo
  {author} {\bibfnamefont {K.-T.}\ \bibnamefont {Ko}}, \bibinfo {author}
  {\bibfnamefont {J.}~\bibnamefont {Seo}}, \bibinfo {author} {\bibfnamefont
  {J.~S.}\ \bibnamefont {Kim}}, \bibinfo {author} {\bibfnamefont {T.-H.}\
  \bibnamefont {Jang}}, \bibinfo {author} {\bibfnamefont {Y.}~\bibnamefont
  {Kim}}, \bibinfo {author} {\bibfnamefont {J.-Y.}\ \bibnamefont {Kim}},
  \bibinfo {author} {\bibfnamefont {S.-W.}\ \bibnamefont {Cheong}}, \ and\
  \bibinfo {author} {\bibfnamefont {J.-H.}\ \bibnamefont {Park}},\ }\href
  {\doibase 10.1103/PhysRevLett.122.207201} {\bibfield  {journal} {\bibinfo
  {journal} {Phys. Rev. Lett.}\ }\textbf {\bibinfo {volume} {122}},\ \bibinfo
  {pages} {207201} (\bibinfo {year} {2019}{\natexlab{b}})}\BibitemShut
  {NoStop}%
\bibitem [{\citenamefont {Yang}\ \emph {et~al.}(2020)\citenamefont {Yang},
  \citenamefont {Fan}, \citenamefont {Wang}, \citenamefont {Khomskii},\ and\
  \citenamefont {Wu}}]{Yang_2019}%
  \BibitemOpen
  \bibfield  {author} {\bibinfo {author} {\bibfnamefont {K.}~\bibnamefont
  {Yang}}, \bibinfo {author} {\bibfnamefont {F.}~\bibnamefont {Fan}}, \bibinfo
  {author} {\bibfnamefont {H.}~\bibnamefont {Wang}}, \bibinfo {author}
  {\bibfnamefont {D.~I.}\ \bibnamefont {Khomskii}}, \ and\ \bibinfo {author}
  {\bibfnamefont {H.}~\bibnamefont {Wu}},\ }\href {\doibase
  10.1103/PhysRevB.101.100402} {\bibfield  {journal} {\bibinfo  {journal}
  {Phys. Rev. B}\ }\textbf {\bibinfo {volume} {101}},\ \bibinfo {pages}
  {100402(R)} (\bibinfo {year} {2020})}\BibitemShut {NoStop}%
\bibitem [{\citenamefont {Blaha}\ \emph {et~al.}()\citenamefont {Blaha},
  \citenamefont {Schwarz}, \citenamefont {Madsen}, \citenamefont {Kvasnicka},\
  and\ \citenamefont {Luitz}}]{WIEN2K}%
  \BibitemOpen
  \bibfield  {author} {\bibinfo {author} {\bibfnamefont {P.}~\bibnamefont
  {Blaha}}, \bibinfo {author} {\bibfnamefont {K.}~\bibnamefont {Schwarz}},
  \bibinfo {author} {\bibfnamefont {G.}~\bibnamefont {Madsen}}, \bibinfo
  {author} {\bibfnamefont {D.}~\bibnamefont {Kvasnicka}}, \ and\ \bibinfo
  {author} {\bibfnamefont {J.}~\bibnamefont {Luitz}},\ }\href@noop {} {\enquote
  {\bibinfo {title} {Wien2k package},}\ }\bibinfo {howpublished}
  {http://www.wien2k.at}\BibitemShut {NoStop}%
\bibitem [{\citenamefont {Anisimov}\ \emph {et~al.}(1993)\citenamefont
  {Anisimov}, \citenamefont {Solovyev}, \citenamefont {Korotin}, \citenamefont
  {Czy\ifmmode~\dot{z}\else \.{z}\fi{}yk},\ and\ \citenamefont
  {Sawatzky}}]{Anisimov_1993}%
  \BibitemOpen
  \bibfield  {author} {\bibinfo {author} {\bibfnamefont {V.~I.}\ \bibnamefont
  {Anisimov}}, \bibinfo {author} {\bibfnamefont {I.~V.}\ \bibnamefont
  {Solovyev}}, \bibinfo {author} {\bibfnamefont {M.~A.}\ \bibnamefont
  {Korotin}}, \bibinfo {author} {\bibfnamefont {M.~T.}\ \bibnamefont
  {Czy\ifmmode~\dot{z}\else \.{z}\fi{}yk}}, \ and\ \bibinfo {author}
  {\bibfnamefont {G.~A.}\ \bibnamefont {Sawatzky}},\ }\href {\doibase
  10.1103/PhysRevB.48.16929} {\bibfield  {journal} {\bibinfo  {journal} {Phys.
  Rev. B}\ }\textbf {\bibinfo {volume} {48}},\ \bibinfo {pages} {16929}
  (\bibinfo {year} {1993})}\BibitemShut {NoStop}%
\bibitem [{\citenamefont {Metropolis}\ and\ \citenamefont
  {Ulam}(1949)}]{Nicholas_1949}%
  \BibitemOpen
  \bibfield  {author} {\bibinfo {author} {\bibfnamefont {N.}~\bibnamefont
  {Metropolis}}\ and\ \bibinfo {author} {\bibfnamefont {S.}~\bibnamefont
  {Ulam}},\ }\href@noop {} {\bibfield  {journal} {\bibinfo  {journal} {J. Am.
  Stat. Assoc.}\ }\textbf {\bibinfo {volume} {44}},\ \bibinfo {pages} {335}
  (\bibinfo {year} {1949})}\BibitemShut {NoStop}%
\bibitem [{\citenamefont {Ou}\ and\ \citenamefont {Wu}(2014)}]{Ou_2014}%
  \BibitemOpen
  \bibfield  {author} {\bibinfo {author} {\bibfnamefont {X.}~\bibnamefont
  {Ou}}\ and\ \bibinfo {author} {\bibfnamefont {H.}~\bibnamefont {Wu}},\ }\href
  {\doibase 10.1038/srep04609} {\bibfield  {journal} {\bibinfo  {journal} {Sci.
  Rep.}\ }\textbf {\bibinfo {volume} {4}},\ \bibinfo {pages} {4609} (\bibinfo
  {year} {2014})}\BibitemShut {NoStop}%
\bibitem [{\citenamefont {Ou}\ \emph {et~al.}(2015)\citenamefont {Ou},
  \citenamefont {Wang}, \citenamefont {Fan}, \citenamefont {Li},\ and\
  \citenamefont {Wu}}]{Ou_2015}%
  \BibitemOpen
  \bibfield  {author} {\bibinfo {author} {\bibfnamefont {X.}~\bibnamefont
  {Ou}}, \bibinfo {author} {\bibfnamefont {H.}~\bibnamefont {Wang}}, \bibinfo
  {author} {\bibfnamefont {F.}~\bibnamefont {Fan}}, \bibinfo {author}
  {\bibfnamefont {Z.}~\bibnamefont {Li}}, \ and\ \bibinfo {author}
  {\bibfnamefont {H.}~\bibnamefont {Wu}},\ }\href {\doibase
  10.1103/PhysRevLett.115.257201} {\bibfield  {journal} {\bibinfo  {journal}
  {Phys. Rev. Lett.}\ }\textbf {\bibinfo {volume} {115}},\ \bibinfo {pages}
  {257201} (\bibinfo {year} {2015})}\BibitemShut {NoStop}%
\bibitem [{\citenamefont {Lee}\ \emph {et~al.}(2010)\citenamefont {Lee},
  \citenamefont {Murray}, \citenamefont {Kong}, \citenamefont {Lundqvist},\
  and\ \citenamefont {Langreth}}]{vdw}%
  \BibitemOpen
  \bibfield  {author} {\bibinfo {author} {\bibfnamefont {K.}~\bibnamefont
  {Lee}}, \bibinfo {author} {\bibfnamefont {E.~D.}\ \bibnamefont {Murray}},
  \bibinfo {author} {\bibfnamefont {L.}~\bibnamefont {Kong}}, \bibinfo {author}
  {\bibfnamefont {B.~I.}\ \bibnamefont {Lundqvist}}, \ and\ \bibinfo {author}
  {\bibfnamefont {D.~C.}\ \bibnamefont {Langreth}},\ }\href {\doibase
  10.1103/PhysRevB.82.081101} {\bibfield  {journal} {\bibinfo  {journal} {Phys.
  Rev. B}\ }\textbf {\bibinfo {volume} {82}},\ \bibinfo {pages} {081101(R)}
  (\bibinfo {year} {2010})}\BibitemShut {NoStop}%
\bibitem [{\citenamefont {Kresse}\ and\ \citenamefont
  {Furthmuller}(1996)}]{vasp}%
  \BibitemOpen
  \bibfield  {author} {\bibinfo {author} {\bibfnamefont {G.}~\bibnamefont
  {Kresse}}\ and\ \bibinfo {author} {\bibfnamefont {J.}~\bibnamefont
  {Furthmuller}},\ }\href {\doibase 10.1103/PhysRevB.54.11169} {\bibfield
  {journal} {\bibinfo  {journal} {Phys. Rev. B}\ }\textbf {\bibinfo {volume}
  {54}},\ \bibinfo {pages} {11169} (\bibinfo {year} {1996})}\BibitemShut
  {NoStop}%
\bibitem [{\citenamefont {Togo}\ and\ \citenamefont {Tanaka}(2015)}]{phonopy}%
  \BibitemOpen
  \bibfield  {author} {\bibinfo {author} {\bibfnamefont {A.}~\bibnamefont
  {Togo}}\ and\ \bibinfo {author} {\bibfnamefont {I.}~\bibnamefont {Tanaka}},\
  }\href@noop {} {\bibfield  {journal} {\bibinfo  {journal} {Scr. Mater.}\
  }\textbf {\bibinfo {volume} {108}},\ \bibinfo {pages} {1} (\bibinfo {year}
  {2015})}\BibitemShut {NoStop}%
\bibitem [{\citenamefont {Baskurt}\ \emph {et~al.}(2020)\citenamefont
  {Baskurt}, \citenamefont {Eren}, \citenamefont {Yagmurcukardes},\ and\
  \citenamefont {Sahin}}]{ref1}%
  \BibitemOpen
  \bibfield  {author} {\bibinfo {author} {\bibfnamefont {M.}~\bibnamefont
  {Baskurt}}, \bibinfo {author} {\bibfnamefont {I.}~\bibnamefont {Eren}},
  \bibinfo {author} {\bibfnamefont {M.}~\bibnamefont {Yagmurcukardes}}, \ and\
  \bibinfo {author} {\bibfnamefont {H.}~\bibnamefont {Sahin}},\ }\href
  {\doibase https://doi.org/10.1016/j.apsusc.2019.144937} {\bibfield  {journal}
  {\bibinfo  {journal} {Appl. Surf. Sci.}\ }\textbf {\bibinfo {volume} {508}},\
  \bibinfo {pages} {144937} (\bibinfo {year} {2020})}\BibitemShut {NoStop}%
\bibitem [{\citenamefont {Iyikanat}\ \emph {et~al.}(2018)\citenamefont
  {Iyikanat}, \citenamefont {Yagmurcukardes}, \citenamefont {Senger},\ and\
  \citenamefont {Sahin}}]{ref2}%
  \BibitemOpen
  \bibfield  {author} {\bibinfo {author} {\bibfnamefont {F.}~\bibnamefont
  {Iyikanat}}, \bibinfo {author} {\bibfnamefont {M.}~\bibnamefont
  {Yagmurcukardes}}, \bibinfo {author} {\bibfnamefont {R.~T.}\ \bibnamefont
  {Senger}}, \ and\ \bibinfo {author} {\bibfnamefont {H.}~\bibnamefont
  {Sahin}},\ }\href {\doibase 10.1039/C7TC05266A} {\bibfield  {journal}
  {\bibinfo  {journal} {J. Mater. Chem. C}\ }\textbf {\bibinfo {volume} {6}},\
  \bibinfo {pages} {2019} (\bibinfo {year} {2018})}\BibitemShut {NoStop}%
\bibitem [{\citenamefont {Bacaksiz}\ \emph {et~al.}(2020)\citenamefont
  {Bacaksiz}, \citenamefont {Yagmurcukardes}, \citenamefont {Peeters},\ and\
  \citenamefont {Milosevic}}]{ref3}%
  \BibitemOpen
  \bibfield  {author} {\bibinfo {author} {\bibfnamefont {C.}~\bibnamefont
  {Bacaksiz}}, \bibinfo {author} {\bibfnamefont {M.}~\bibnamefont
  {Yagmurcukardes}}, \bibinfo {author} {\bibfnamefont {F.~M.}\ \bibnamefont
  {Peeters}}, \ and\ \bibinfo {author} {\bibfnamefont {M.~V.}\ \bibnamefont
  {Milosevic}},\ }\href {\doibase 10.1088/2053-1583/ab6d79} {\bibfield
  {journal} {\bibinfo  {journal} {2D Mater.}\ }\textbf {\bibinfo {volume}
  {7}},\ \bibinfo {pages} {025029} (\bibinfo {year} {2020})}\BibitemShut
  {NoStop}%
\bibitem [{\citenamefont {Khomskii}(2014)}]{Khomskii_2001}%
  \BibitemOpen
  \bibfield  {author} {\bibinfo {author} {\bibfnamefont {D.~I.}\ \bibnamefont
  {Khomskii}},\ }\href@noop {} {\emph {\bibinfo {title} {Transition Metal
  Compounds}}}\ (\bibinfo  {publisher} {Cambridge University Press},\ \bibinfo
  {year} {2014})\BibitemShut {NoStop}%
\end{thebibliography}%

\end{document}